\documentclass{appolb}
\usepackage{cite,graphicx,amsmath,amssymb}
\newcommand{\sla}[1]{\ifmmode%
  \setbox0=\hbox{$#1$}%
  \setbox1=\hbox to\wd0{\hss$/$\hss}\else%
  \setbox0=\hbox{#1}%
  \setbox1=\hbox to\wd0{\hss/\hss}\fi%
  #1\hskip-\wd0\box1 }

\begin{document}
\title{Accurate backgrounds to Higgs production at the LHC
\thanks{Presented at the ``Physics at LHC'' Conference, Cracow, Poland, July 3--8, 2006.}%
}
\author{Nikolas Kauer
\address{Institut f\"ur Theoretische Physik und Astrophysik,\\
Universit\"at W\"urzburg, D-97074 W\"urzburg, Germany}
}
\maketitle
\begin{abstract}
Corrections of 10-30\% for backgrounds to the $H\to WW\to \ell^+\ell^-\sla{p}_T$ search
in vector boson and gluon fusion 
at the
LHC are reviewed to make the case for precise and accurate theoretical background 
predictions.
\end{abstract}
  
\section{Introduction}

With full energy collisions in the Large Hadron Collider (LHC) scheduled for 
spring 2008, in the near future the TeV scale will become directly accessible in 
experiments.  Theoretical arguments and precision measurements 
indicate that ground-breaking discoveries regarding the mechanism
of electroweak symmetry breaking and the origin of mass can be expected
\cite{nk_higgsphysics_tutorials}.
The discovery and analysis of the Higgs boson predicted by the Standard Model (SM)
with Higgs mechanism \cite{nk_higgsmechanism} or additional particles, for instance 
in supersymmetric extensions \cite{nk_susy}, are the primary goal of the LHC physics 
program \cite{nk_higgsphysics_reviews}.

For that purpose theoretical predictions are needed not only for signal processes,
but also for background processes in search channels that do not allow for
background determination from data (\eg via sideband interpolation).  
While most background processes have previously been considered as signals, these calculations frequently employ approximations that are appropriate 
for signal selections, but may lead to inaccurate predictions for suppressed 
backgrounds.
In many cases, higher order corrections have been 
calculated for inclusive cross sections, but not
for fully differential cross sections due to the increased 
complexity.  While inclusive corrections are in good 
approximation applicable to signal predictions, only fully differential
corrections
are guaranteed to provide accurate predictions for suppressed backgrounds.
Although the discovery significance $S/\sqrt{B}$ is less affected by background 
corrections compared to signal corrections of similar size and sign,
the applied selection cuts can strongly enhance suppressed 
background $K$-factors, but will not have a significant effect on the signal 
$K$-factor.

\begin{figure}[htb]
\vspace{0.3cm}
\begin{center}
\includegraphics[width=8.0cm]{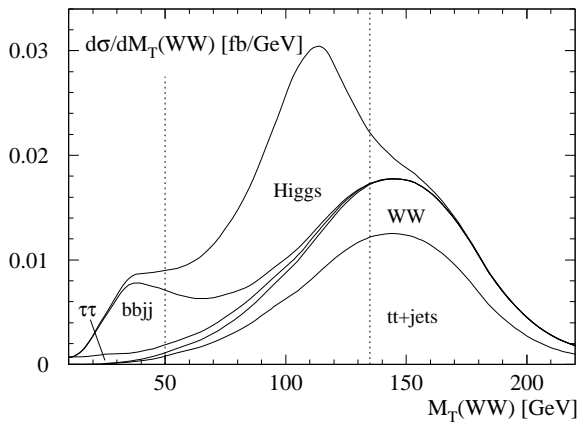} \\
\end{center}
\caption{Distribution of the $WW$ ``transverse mass'' $d\sigma/dM_T$ 
as Higgs invariant mass proxy ($M_H$ = 115 GeV) for $e^\pm\mu^\mp\sla{p}_T$ events 
in vector boson fusion at the LHC.
The areas between curves represent the contributions from the signal and 
the various background classes, as indicated.
For further details see Ref.~\protect\cite{Kauer:2000hi} (Fig.~4).}
\vspace{0.5cm}
\label{fig:vbf_mT}
\end{figure}

In the following 
we focus on the production of a SM Higgs boson in vector boson or gluon fusion that 
decays into $W$ bosons, which in turn decay leptonically.  These search channels 
contribute significantly to the LHC SM Higgs discovery potential for Higgs masses 
between 120 and 200 GeV \cite{nk_vbf_hwwllvv,Kauer:2000hi,nk_gf_hwwllvv}.
Since the decay neutrinos escape detection, the Higgs 
momentum cannot be reconstructed.  Signal and background can thus not be separated 
experimentally.  Higgs observation becomes a counting experiment and reliable 
background predictions essential.  For vector boson fusion (see Fig.~\ref{fig:vbf_mT}) 
as well 
as gluon fusion, the dominant irreducible background is $W$-pair production and the 
leading reducible background is $t\bar{t}$ (+ jets) production.

\section{Top background corrections}

Perturbative predictions for the production and decay of heavy particles
have frequently relied on the narrow-width approximation (NWA), which significantly 
reduces the number of Feynman diagrams that have to be taken into account.
The parametric error estimate of ${\cal O}(\Gamma/m)\sim 1\%$ is reliable
if applied selection cuts do not affect the resonant region and additional
sub- and nonresonant diagrams with the same final state can be neglected.  
While this is usually the case for SM signal cross sections, the same is not
true for suppressed background cross sections.
For backgrounds, it is thus important to verify that the NWA is appropriate by
comparing with calculations that take into account the complete fixed-order amplitude.
For the Higgs search channels considered here, this comparison has been performed for 
the dominant reducible background due to top pair production in Ref.~\cite{nk_topcorr}.  
Off-shell predictions with the complete leading-order (LO) amplitude for $t\bar{t}$, 
$t\bar{t}$ + 1 jet and
$t\bar{t}$ + 2 jets (with 87, 600 and 5820 Feynman diagrams contributing for the at 
the LHC dominant $gg$ subprocess, respectively) were obtained using the 
\texttt{PP2TTNJ} program \cite{hepsource_programs}.\footnote{
For $t\bar{t}$ production without additional jets, the subset of all graphs that 
contain an intermediate $W^+$ and $W^-$ boson is available in 
the user process package \texttt{AcerMC} \protect\cite{nk_acermc}, 
which features interfaces 
to the general-purpose event generators PYTHIA and HERWIG.}
The complete LO predictions are typically 10-20\% larger than the
corresponding NWA predictions.  The standard NWA uncertainty estimate
is thus not applicable.  Note that this even holds for the total cross section 
at the LHC, which is enhanced by about 5\% if single- and nonresonant diagrams 
are taken into account.\footnote{The NWA uncertainty of the Tevatron total cross 
section, on the other hand, is estimated correctly.}

In gluon fusion the applied central jet veto increases the single-resonant
contribution\footnote{with respect to top} to approximately match 
the double-resonant contribution.
Since the LO background uncertainty is large and the techniques for a complete 
$WWb\bar{b}$ calculation at next-to-leading order (NLO) are still under development (see Sec.~\ref{extrapol}),
inspired by the natural separation of the double- and  single-resonant graphs  
into gauge-invariant subsets ($t\bar{t}$ and $Wt$) 
at LO when the NWA is applied, the authors of Ref.~\cite{Campbell:2005bb}
studied approaches to combine 
the corresponding $t\bar{t}$ and $Wt$ cross sections known at NLO and 
discuss heuristic prescriptions to handle the overlap between 
real corrections to $Wt$ and $t\bar{t}$ at LO.

\section{$WW$ background corrections}

\begin{figure}[tb]
\vspace{0.cm}
\begin{minipage}[c]{.49\linewidth}
\flushright \includegraphics[height=5.8cm, clip=true, angle=90]{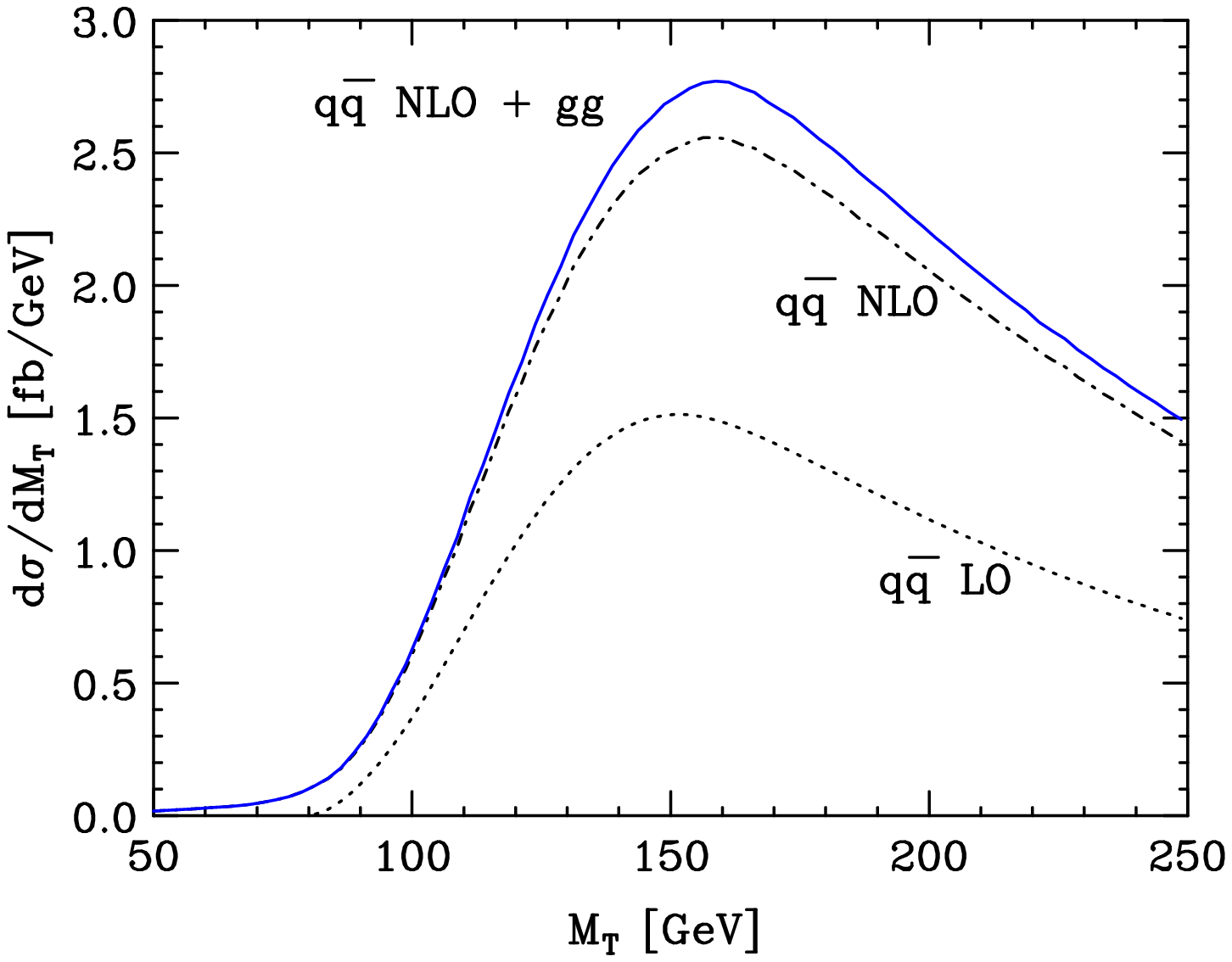}
\end{minipage} \hfill
\begin{minipage}[c]{.49\linewidth}
\flushleft \includegraphics[height=5.8cm, angle=90]{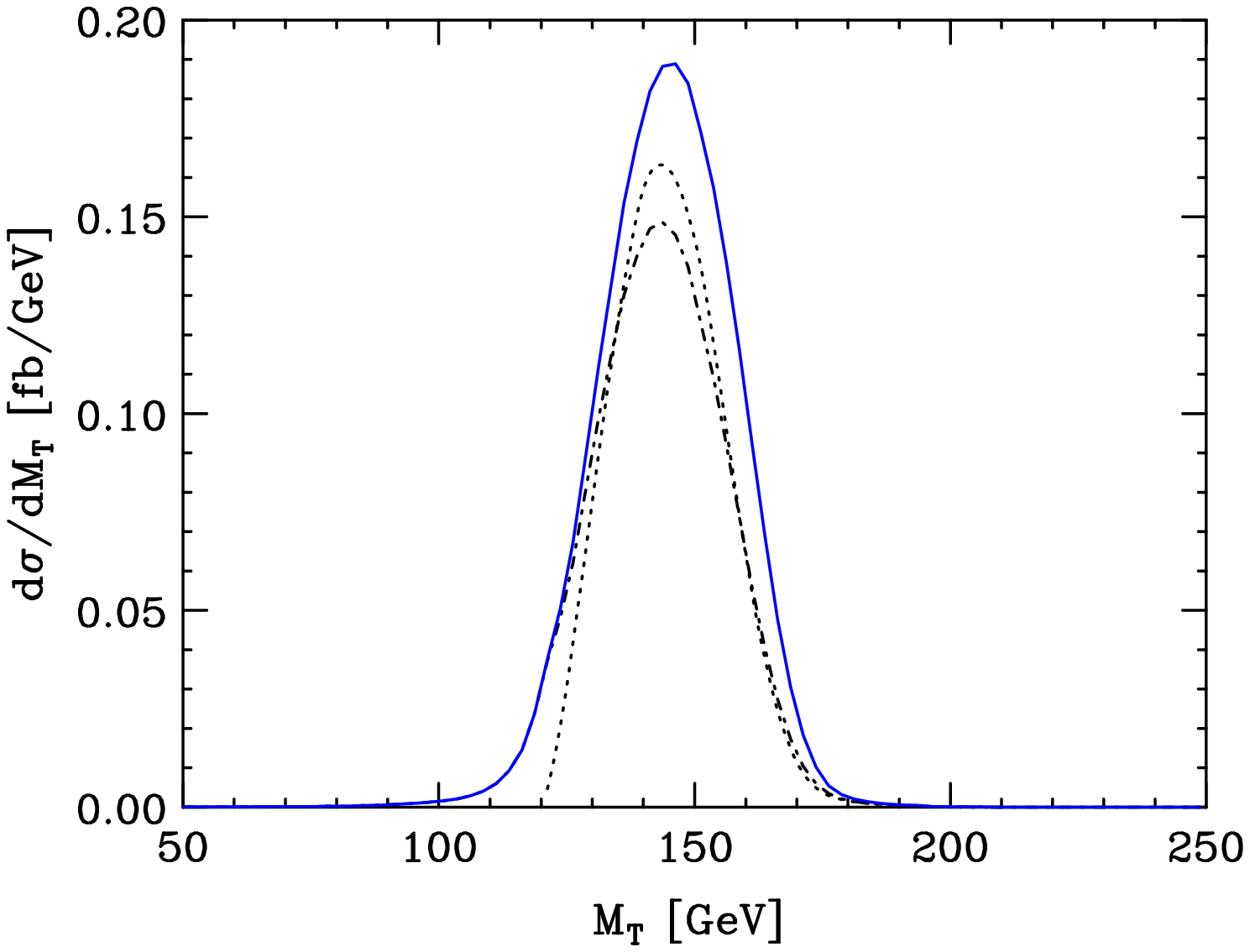}
\end{minipage}\\[0.2cm]
\caption{{\label{fig:mT_ggWW} Distribution of $M_T$ (see
    Fig.~\protect\ref{fig:vbf_mT}) with standard cuts$^\text{\protect\ref{nk_cuts}}$ (left) and Higgs search cuts$^\text{\protect\ref{nk_cuts}}$ (right).  Displayed are the total background from quark scattering
    at NLO and gluon-fusion (solid), and from quark scattering alone
    at LO (dotted) and NLO (dot-dashed).}}
\vspace{0.5cm}
\label{fig:gg_mT}
\end{figure}

The dominant irreducible background to $pp \to H \to
WW \to \ell^+\ell^-\sla{p}_T$ is $W$-pair production, which 
occurs in quark-antiquark annihilation.  The gluon-induced subprocess 
formally enters at next-to-next-to-leading order (NNLO), but its importance  
is enhanced by the large gluon flux at the LHC and by experimental Higgs 
search cuts.\footnote{Standard cuts are $p_{T,\ell} > 20 \text{ GeV}, |\eta_\ell| < 2.5, \sla{p}_T > 25 \text{ GeV}$.  Additional Higgs search cuts are $\Delta\phi_{T,\ell\ell} < 45^\circ$, $m_{\ell\ell} < 35\ \text{GeV}$, $\text{jet veto:}$ $p_{Tj} > 20\ \text{GeV}\ \text{and}\ |\eta_j| < 3$, $35\ \text{GeV} <  p_{T\ell,\text{max}} < 50\ \text{GeV}, 25\ \text{GeV} <  p_{T\ell,\text{min}}$.\label{nk_cuts}}
Recently, a fully 
differential calculation of $gg \to W^{\ast}W^{\ast} \to
\ell\bar{\nu}\bar{\ell'}\nu'$ including the top-bottom massive quark loop
contribution and the intermediate Higgs contribution with full 
spin and decay angle correlations and allowing for arbitrary
invariant masses of the $W$ bosons has been completed \cite{nk_ggww_papers}.  
The gluon-induced contribution enhances the NLO $WW$ background prediction 
by approximately $30\%$.  Though NNLO, it is the dominant higher-order correction
(see Fig.~\ref{fig:gg_mT}).  
Signal-background interference effects range from 
--4 to +11\% for intermediate Higgs masses when Higgs search cuts are applied
(see Table \ref{tbl:interference}).
The calculation is available as event generator \texttt{GG2WW} \cite{hepsource_programs}.  Work on a program \texttt{GG2ZZ} \cite{hepsource_programs}
for the process $gg\to Z^\ast Z^\ast \to$ 4 charged leptons is in progress.

\begin{table}[tb]
\caption{\label{tbl:interference} 
  Interference effects between the signal and gluon-induced background
  processes $gg (\to H) \to W^{\ast}W^{\ast}\to
   \ell\bar{\nu}\bar{\ell'}\nu'$ with Higgs search selection cuts (see Ftn.~\protect\ref{nk_cuts})
  and cross sections in fb.}
\vspace*{.2cm}
\begin{center}
\renewcommand{\arraystretch}{1.5}
\begin{tabular}{|c|c|c|c|}
\hline
\multicolumn{1}{|c|}{$M_H {\rm [GeV]}$} &
$140$ &
$170$ &
$200$ \\
\hline
$\sigma[{\rm signal}]$  & 1.8852(5) & 12.974(2) & 1.6663(7) \\
 \hline
$\sigma[{\rm bkg(gg)}]$ & \multicolumn{3}{c|}{1.4153(3)} \\
 \hline
$\sigma[{\rm signal+bkg(gg)}]$ & 3.174(2) & 15.287(6) & 3.413(2)  \\
 \hline
$\frac{\sigma[{\rm signal+bkg(gg)}]}{\sigma[{\rm signal}]+\sigma[{\rm bkg(gg)}]}$
& $0.962$ & $1.062$ & $1.108$ \\[1ex]
 \hline
\end{tabular}
\end{center}
\vspace*{.0cm}
\end{table}

\section{BSM corrections}

\begin{figure}[htb]
\vspace{0.3cm}
\begin{minipage}[c]{.49\linewidth}
\flushright \includegraphics[height=5.5cm, clip=true, angle=90]{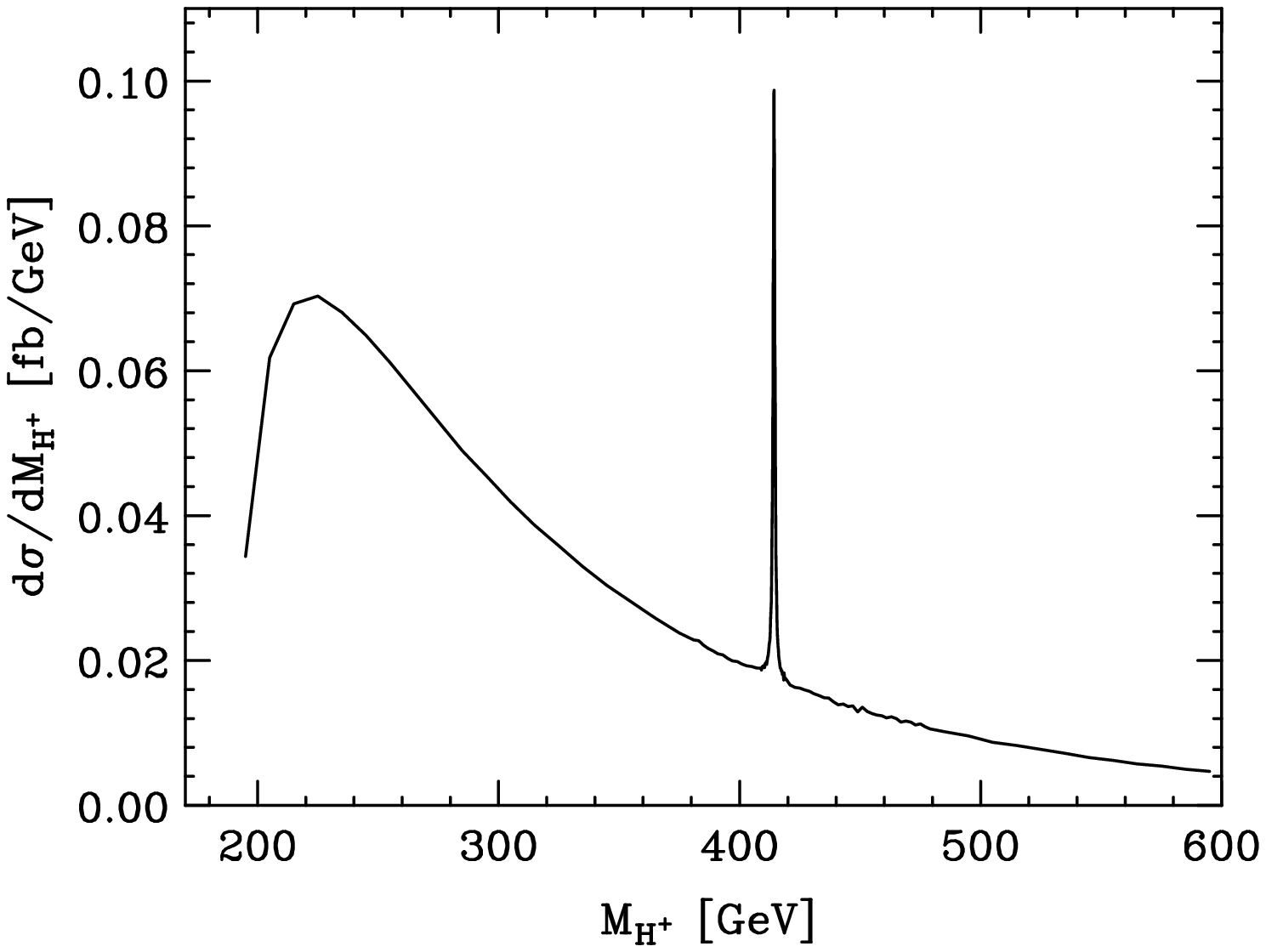}
\end{minipage} \hfill
\begin{minipage}[c]{.49\linewidth}
\flushleft \includegraphics[height=5.5cm, clip=true, angle=90]{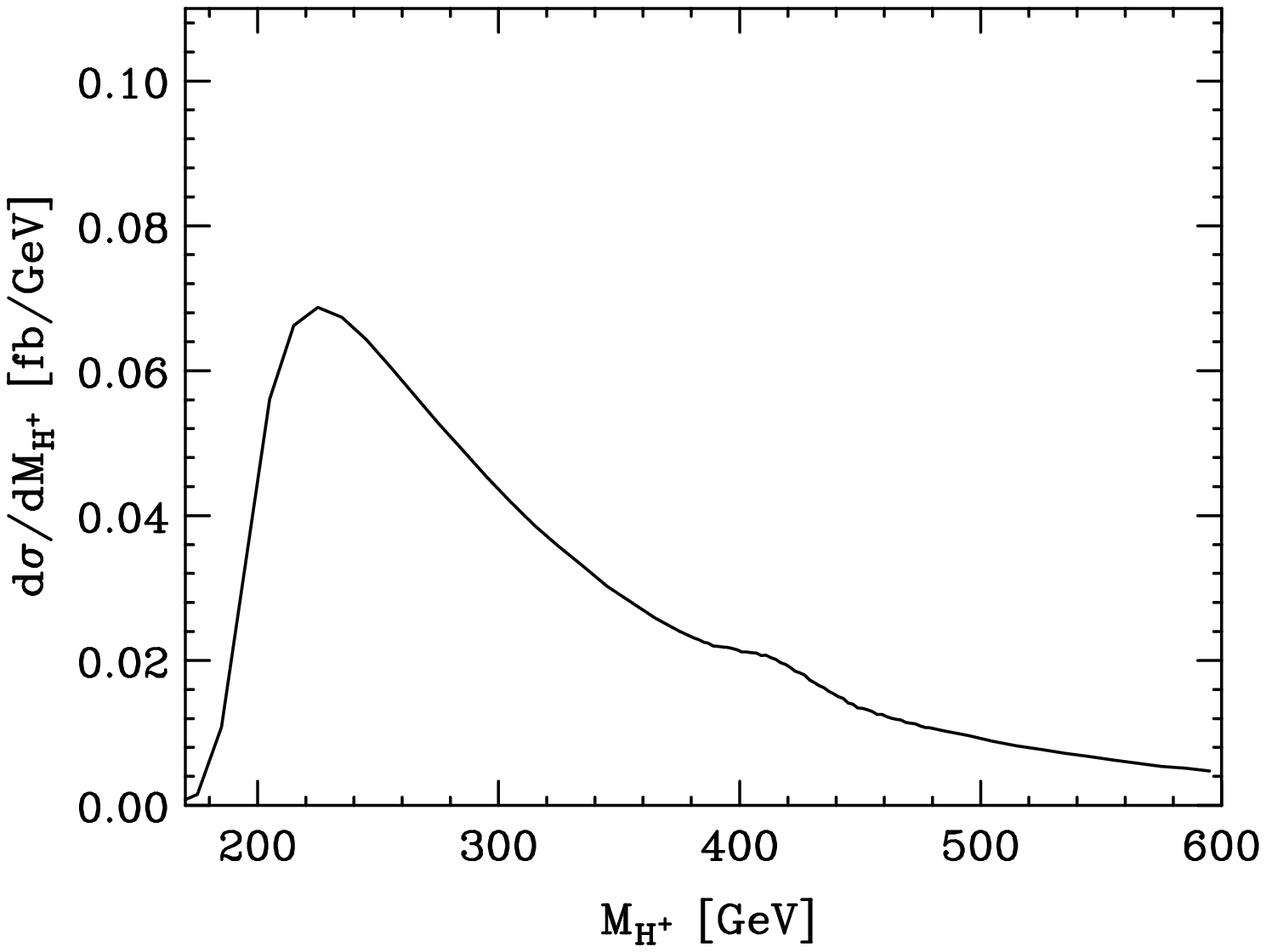}
\end{minipage}\\[0.2cm]
\caption{\label{fig:hcinvmass}
$H^+$ invariant mass distribution for the process 
$g\bar{b}\to (H^+\to h\,W^+)\:\bar{t}$ at the LHC calculated with complete LO amplitude
with perfect detector resolution
(left) and with realistic detector resolution (right) at SPS1a.  No selection cuts
are applied.
}
\end{figure}

In extensions of the SM additional heavy particles lead to more complicated
decay chains.  To simplify calculations in these models, the NWA has been applied 
extensively, even though the number of nonresonant diagrams that are neglected 
increases.  Consider, for instance, $g\bar{b}\to (H^+\to h\,W^+)\:\bar{t}$,
which has been studied in NWA in Ref.~\cite{Drees:1999sb}.
For SPS1a with no selection cuts applied one obtains at the LHC 
$\sigma_\text{complete}/\sigma_\text{NWA}$ = 110 (see Fig.~\ref{fig:hcinvmass}).
The 8 nonresonant diagrams neglected in NWA dominate the cross section.\footnote{
Interference effects are negligible.}  This dramatic effect serves as a warning to 
expect surprises.\footnote{Note that the branching ratio is small in the MSSM. For SPS1a, $\text{BR}_{H^+\to hW^+}$ = $2\cdot 10^{-3}$.  It could, however, be much larger in the NMSSM or
general 2HDMs.}
Other examples and a more detailed discussion can be found in Ref.~\cite{nk_nwa_bsm}.

\section{Reducing the theoretical uncertainty\label{extrapol}}

\begin{figure}[htb]
\vspace{0.3cm}
\begin{center}
\begin{minipage}[c]{.49\linewidth}
\flushright \includegraphics[width=4.8cm, angle=90]{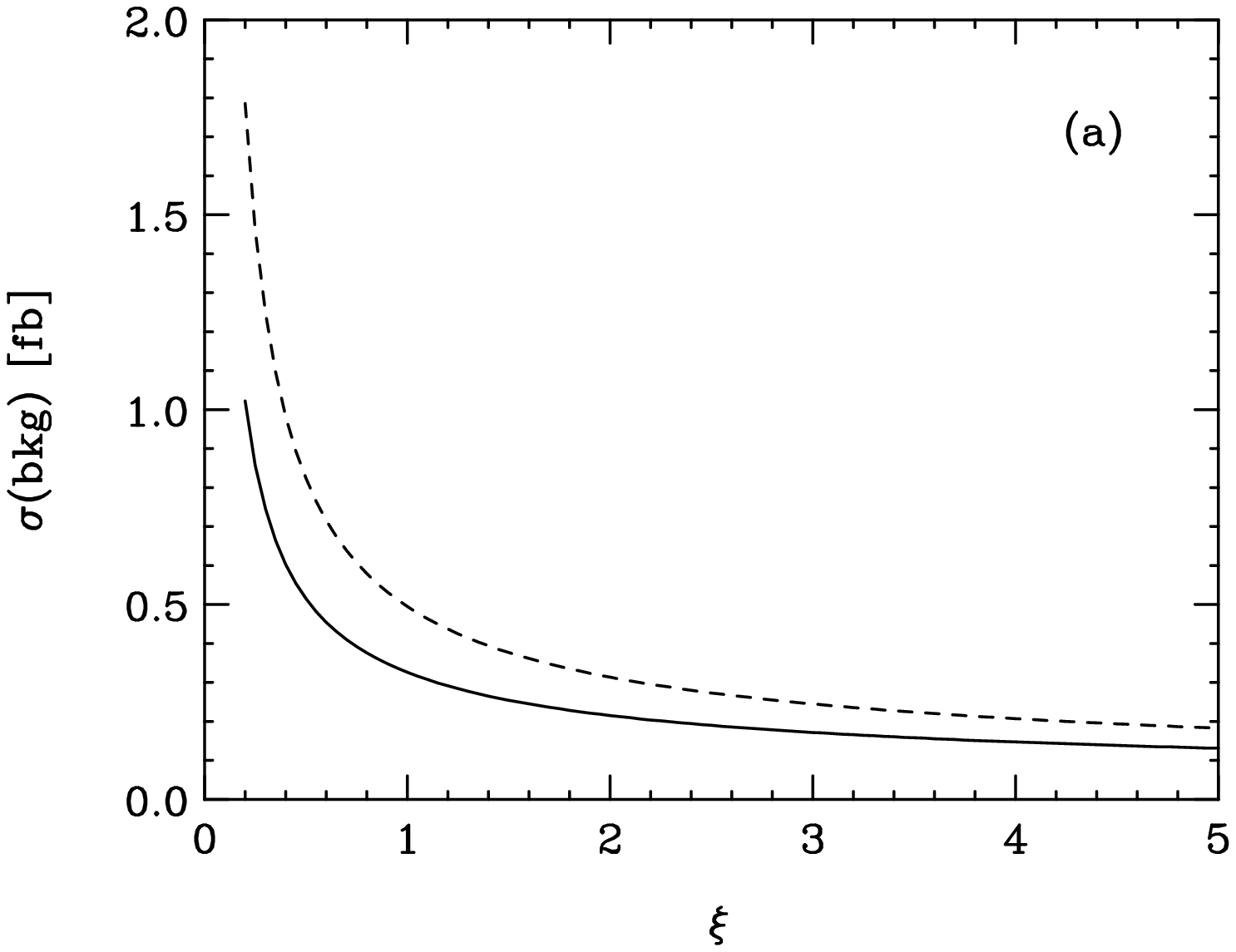}
\end{minipage} \hfill
\begin{minipage}[c]{.49\linewidth}
\flushleft \includegraphics[width=4.8cm, angle=90]{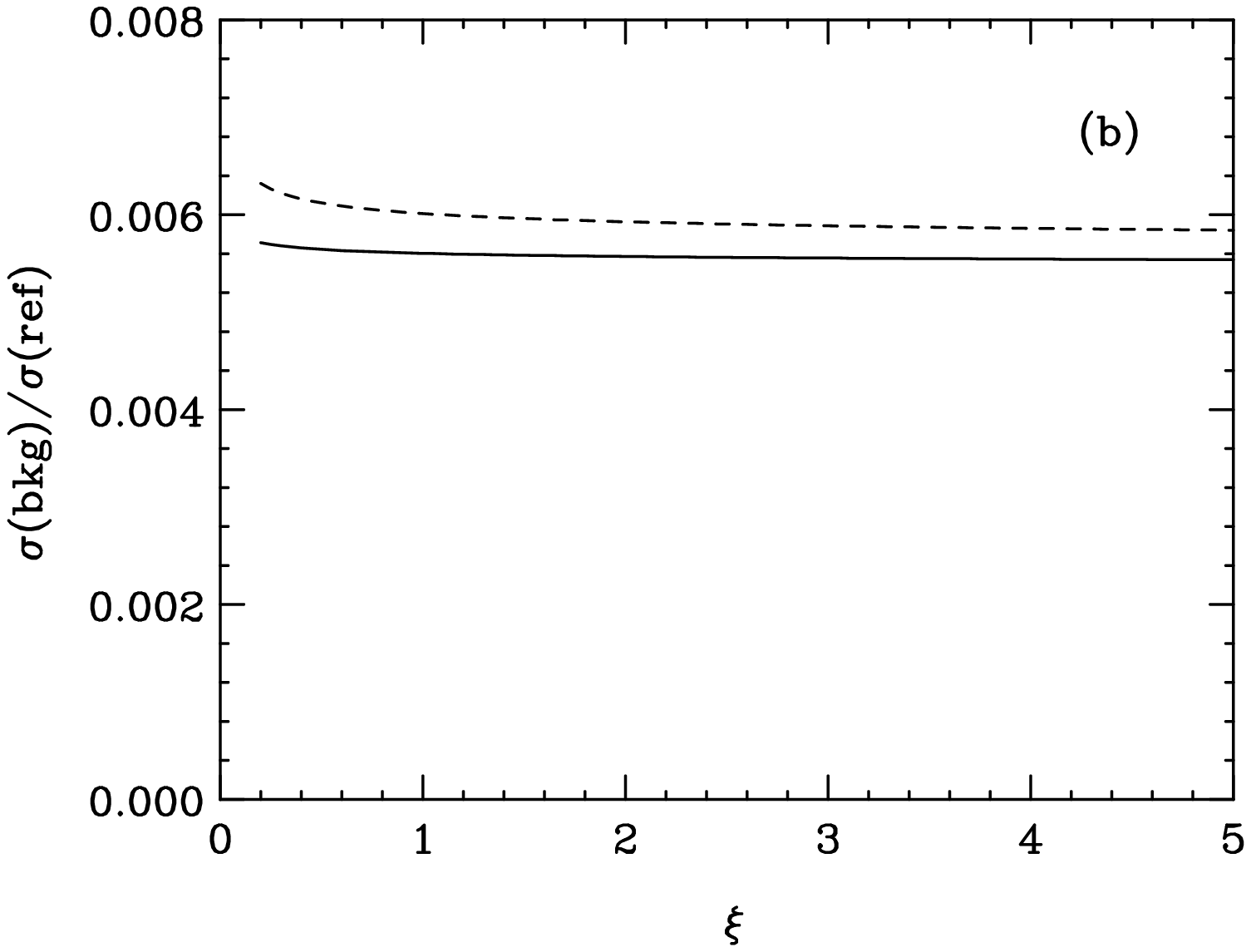} 
\end{minipage}
\caption{\label{fig:vbf-scalevariation}
  Renormalization and factorization scale variation of LO $t\bar{t}j$ background cross
  section (a) and ratio with reference cross section (b) to the
  $H\to WW\to \ell^+ \ell^-\sla{p}_T$ search in vector boson fusion at the LHC
  for different scale definitions.  For details see Ref.~\protect\cite{Kauer:2004fg} (Fig.~1).
}
\end{center}
\end{figure}

LO QCD cross sections are affected by large scale uncertainties (see Fig.~\ref{fig:vbf-scalevariation}(a)).  The following complementary approaches allow to obtain theoretical predictions with an uncertainty of $\lesssim$ 10\%.
\begin{itemize}
\item calculation of higher order corrections
\item extrapolation of measured cross sections
\end{itemize}
Since exact cross sections are scale independent, higher fixed-order results tend to 
have a strongly reduced scale uncertainty.  In special kinematic regimes large logarithms occur and need to be resummed in order to obtain an acceptable convergence of the
perturbative series.  Without NWA decomposition, calculations of backgrounds usually
involve many-particle final states, and the calculational complexity increases 
considerably order by order.  Adequate methods that allow to efficiently evaluate  
and integrate the 1-loop multi-leg amplitudes of 2 $\to$ 3 and 2 $\to$ 4 processes at 
NLO are therefore currently under development (see \eg \cite{nk_golem}).

In ratios the scale dependence of fixed-order cross sections can partly compensate 
resulting in 
a significantly 
reduced theoretical uncertainty (see Fig.~\ref{fig:vbf-scalevariation}(b)).  
A reference cross section that can be
measured with low uncertainty can thus be extrapolated to the background region
without introducing large theoretical uncertainties \cite{Eboli:2000ze,Kauer:2004fg}:
\[
\sigma_{bkg} \quad \approx \quad \underbrace{\left(
     \frac{\sigma_{bkg,\;\text{fixed-order prediction}}}{\sigma_{ref,\;\text{fixed-order prediction}}}\right)}_{\text{low theoretical uncertainty}}
     \quad\times\quad
\underbrace{\sigma_{ref,\;\text{measured}}}_{\text{low experimental uncertainty}}\ \ .
\]\\[-0.05cm]

\section{Summary}
Theoretical arguments and precision measurements 
indicate that ground-breaking discoveries regarding the mechanism
of electroweak symmetry breaking and the origin of mass
can be expected at the LHC.
For that purpose theoretical predictions are needed not only for signal processes,
but also for important background processes in search channels that do not allow for
background determination from data.
We reviewed NWA and higher order corrections of 10-30\% for the dominant 
$t\bar{t}$ (+ jets) and $WW$ backgrounds 
to the Higgs boson search channel $H\to WW\to \ell^+\ell^-\sla{p}_T$ in 
vector boson and gluon fusion.  In 
extensions of the SM, where additional heavy particles lead to more complicated
decay chains and enhanced nonresonant contributions,   
NWA corrections can be 
significant not only for suppressed backgrounds, but also for signal cross sections.
Two complementary approaches can be applied to reduce the theoretical uncertainty of 
fixed-order background predictions.  
We conclude that enhanced background predictions are necessary and 
feasible and that improved parton-level programs and event generators are 
available \cite{hepsource_programs}.\\[0.2cm]

The author thanks all collaborators and contributors.
The work presented was supported by 
the German Bundesministerium f\"ur Bildung und Forschung  
under contract number 05HT1WWA2, the Deutsche Forschungsgemeinschaft (DFG) 
under contract number BI 1050, the DFG Sonderforschungsbereich/Transregio 9 
``Computer-gest\"{u}tzte Theoretische Teilchenphysik'', the British 
Particle Physics and Astronomy Research Council under grant 
numbers PPA/G/O/2000/00456 and PPA/G/O/2002/00465 and the 
U.~S.~Department of Energy under contract number DE-FG02-95ER40896.

\end{document}